# Sierpinski Gaskets for Logic Functions Representation

Denis V. Popel
Department of Computer Science,
Baker University, Baldwin City,
KS 66006-0065, U.S.A.
Denis.Popel@bakeru.edu

Anita Dani
Department of Computer Science,
University of Wollongong (Australia),
Dubai Campus, U.A.E.,
AnitaDani@uowdubai.ac.ae

## Abstract

*This paper introduces a new approach to represent logic functions in the form of Sierpinski Gaskets. The structure of the gasket allows to manipulate with the corresponding logic expression using recursive essence of fractals. Thus, the Sierpinski gasket's pattern has myriad useful properties which can enhance practical features of other graphic representations like decision diagrams. We have covered possible applications of Sierpinski gaskets in logic design and justified our assumptions in logic function minimization (both Boolean and multiple-valued cases). The experimental results on benchmarks with advances in the novel structure are considered as well.*

## 1 Introduction

Over the years, many important problems in digital circuits synthesis have been approached using graph-based data structures: decision trees, decision diagrams ... etc. The superior example is *Binary Decision Diagram* (BDD) which has become the advanced structure in VLSI CAD for representation and manipulations of logic functions [1]. The applications of BDD techniques cover practically all stages of digital circuits design from the initial representation of Boolean functions [7] to the synthesis of the desirable circuit [5]. But still originally introduced BDDs cannot be directly applied to synthesize circuits derived from AND/EXOR expressions. Such circuits are demanded by VLSI CAD, because of economical implementation (in terms of gates and interconnections) and high testability. This is particularly efficient for error control and arithmetic circuits, encrypting and coding schemes. In this paper, we address the problem of representation of the general class of AND/EXOR expressions, named *Exclusive Sum-Of-Products* (ESOP). Following the above problem, we introduce another graph-based structure popular in *fractal* theory to represent ESOP expressions and their multiple-valued counterparts.

One is familiar with the useful properties of *Pascal triangle* which represents binomial coefficients. Another advanced mathematical structure, which is used frequently in fractal theory, is *Sierpinski gasket* [6]. It should be mentioned that Sierpinski gasket is Pascal triangle modulo two – the EXOR operation is used instead of addition [3]. Sierpinski gasket also referred to as Sierpinski triangle has a number of interesting properties that W.Sierpinski discussed in his paper [9]. Areas of application of Sierpinski fractals are graphic design (Figure 1 shows (a) Sierpinski gasket and (b) Sierpinski pyramid as the samples of fractals implementation) [6], telecommunication (the antenna of cellular phone has fractal square structure), tautology mapping [11], symmetry handling in logic design [2, 12] ... etc. In this paper, we propose a new approach to utilize the structure of Sierpinski gasket to represent ESOP expressions. Preliminary case study results have been published in [3]. We also generalize this approach to multiple-valued functions.

The rest of the paper is organized as follows. In Section 2, we collect necessary definitions and provide basic terminology. Section 3 introduces the structure of Sierpinski gasket and describes the techniques of constructing Sierpinski gasket using different function descriptions. The correspondence between Sierpinski gaskets and logic expressions is outlined in this Section. Section 4 is dedicated to a minimization algorithm and experimental results on Boolean and multiple-valued functions. Finally, Section 5 concludes the paper.

## 2 Basic Concepts

We consider a *multiple-valued function* $f: \mathbf{M}^n \to \mathbf{M}$ over the variable set $X = \{x_1, \cdots, x_n\}$, where $\mathbf{M}=\{0, 1, \ldots, r-1\}$, $n$ is the number of $r$-valued variables. $f_{|x_i=v} = f(x_1, \ldots, x_{i-1}, v, x_{i+1}, \ldots, x_n)$ is called a *cofactor* of $f$, when $x_i$ takes value $v \in \{0, 1, \ldots r-1\}$.

In the Galois field representation of MVL functions, $p$ complements of a $p-$valued variable are usually consid-ered. They are defined by $^{i-}x = x+i$, $i = 1,\ldots,p-1$, where '+' denotes the addition in the considered Galois field. Therefore, for functions over GF(4), we consider four literals $x$, $^{1-}x$, $^{2-}x$, $^{3-}x$ for each variable $x$.

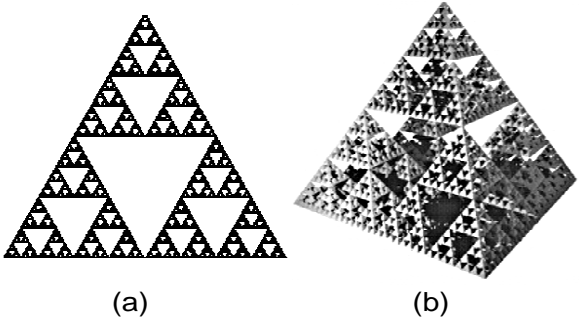

**Figure 1. Sierpinski gasket and pyramid**

For simplicity, the presentation in this paper is restricted to Boolean and 4-valued functions. However, our method can be applied to logic functions of any radix. Let us review the representation forms of logic functions.

## 2.1 Boolean functions

It has been shown in [8] that restricted classes of AND/EXOR expressions, like Fixed Polarity Reed-Muller (FPRM) expressions, can be represented efficiently.

**Definition 1.** *FPRM is an exclusive OR of AND product terms, where each variable appears either complemented or uncomplemented, but not both.*

Based on Sierpinski gasket we are able to represent general class of AND/EXOR expressions – ESOP.

**Definition 2.** *ESOP is EXOR sum of product terms with complemented and uncomplemented variables.*

Observe that in ESOP expression a variable may appear in a product term in complemented or uncomplemented forms. In case of ESOP, the polarity vector $[c_0 c_1 \ldots c_n]$ is associated with given function $f$, where the entry $c_i = \{0, 1, d\}$ is 0 if $x_i$ appears always in uncomplemented form, the entry is 1 for the variable in complemented form and is $d$ if $x_i$ is used in complemented as well as uncomplemented forms (mixed mode) [4]. Note that different polarity vectors for the same function cause different cost characteristics of AND/EXOR expressions (number of product terms, number of literals) [5].

**Example 1.** *For the function $f = x_1 \vee \overline{x}_3 \cdot \overline{x}_2$ with the truth vector* [10001111], *FPRM with polarity vector* [111] *is $f = 1 \oplus \overline{x}_1 \oplus \overline{x}_1\overline{x}_2\overline{x}_3$ (3 product terms), and ESOP with polarity vector* [d11] *is $f = x_1 \oplus \overline{x}_1\overline{x}_2\overline{x}_3$ (2 product terms). The effect in number of product terms has been obtained for different polarity vectors.*

## 2.2 Multiple-valued functions

The fixed polarity expansion for 4-valued function can be introduced.

**Definition 3.** *(D. Green [4]) Reed-Muller expansion of 4-valued function takes the form*

$$f = c_0 + c_1 x + c_2 x^2 + c_3 x^3,$$

*where operations are fulfilled in GF(4).*

In our approach, we extend the representation of 4-valued functions using mixed polarities:

$$f = c_0 + c_1\,^{1-}x + c_2\,^{1-}x^2 + c_3\,^{1-}x^3,$$
$$f = c_0 + c_1\,^{2-}x + c_2\,^{2-}x^2 + c_3\,^{2-}x^3,$$
$$f = c_0 + c_1\,^{3-}x + c_2\,^{3-}x^2 + c_3\,^{3-}x^3.$$

All possible combinations of four expansions specified above can be represented using Sierpinski pyramid.

## 3 Sierpinski Gasket

**Definition 4.** *Sierpinski gasket in our study is a recursively connected graph with the vertex set and the edge set, where:*

**(i)** *Each vertex is labeled by a product term $p$ assigned, as a decision term. Each vertex associated with product term $p$ has at least two connected vertexes $p_1$ and $p_2$ correspondingly which are related using the following triangle rule:*

$$p \oplus p_1 \oplus p_2 = 0. \quad (1)$$

**(ii)** *Each triangle contains three sub-triangles except the primitive one. The biggest equilateral triangle in Sierpinski gasket encloses three equilateral triangles with the base one third of the size of the original, and with hole in the center. Replacement of each of those sub-triangles gives three more sub-triangles, to obtain the gasket as nine triangles, each with a base of length 1/9. The repeating of this replacement $n-1$ times produces the final gasket.*

The vertexes of Sierpinski gasket are addressed according to a *coordinate convention* (Section 3.1).

**Example 2.** *The gasket for the function $f$ from Example 1 is given in Figure 2.*

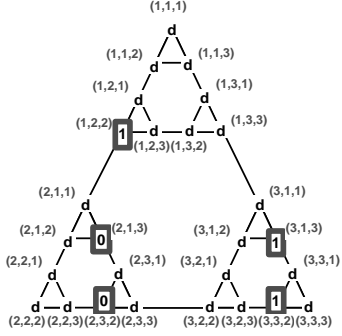

**Figure 2. Sierpinski gasket for Example 1**

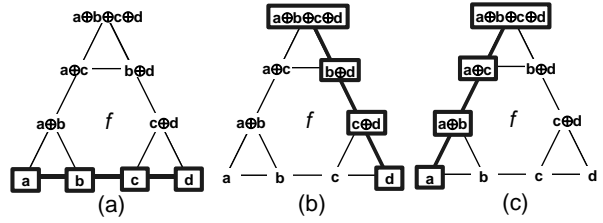

**Figure 3. Sierpinski gasket and the coefficients of positive polarity and negative polarity Reed-Muller expressions**

As well as other fractals, Sierpinski gasket can be defined as a recursive procedure. We propose to use the following recursive procedure to build Sierpinski Gasket ($SG$) for the function $f$ (see Definition 4 for details):

$$SG(f) = SG(f_{|x=0}) \cup SG(f_{|x=1}) \cup SG(f_{|x=d})$$

Another technique for building Sierpinski gasket is to use the truth vector of the function $f$:

**Step 1.** Start building by arranging elements of the truth vector at the base level ($level = 0$).

**Step 2.** Select all pairs of elements, starting from the left side, EXOR them, and than write them down at the upper level.

**Step 3.** Repeat the second step until the level contains one element only.

**Example 3.** *Consider a two variables function $f$ specified by truth vector $[abcd]$. Figure 3(a) shows the base level (truth vector) for constructing Sierpinski gasket. The coefficients $\{a \oplus b \oplus c \oplus d, b \oplus d, c \oplus d, d\}$ of positive polarity Reed-Muller expression for given function are represented by the right side of the triangle (Figure 3(b)), coefficients $\{a \oplus b \oplus c \oplus d, a \oplus c, a \oplus b, a\}$ of negative polarity Reed-Muller expression are located on the left side of the triangle (Figure 3(c)).*

### 3.1 Properties

To manipulate with Sierpinski gasket as regular fractal structure, we need to apply a coordinate convention. Two options can be analyzed.

**Coordinate convention 1.** Each element (vertex) of the Sierpinski gasket can be addressed as two-dimentional array $(i, j)$, where the first $i$ index corresponds to the binary code of the product term of variables without negation, and the second $j$ is equivalent to the binary code of the product term with complemented variables.

**Lemma 1.** *The sum of both coordinates $i$ and $j$ is equal to level number of the Sierpinski gasket.*

**Theorem 1.** *The total memory size (in elements) that is needed to represent a function $f$ using Sierpinski gasket and the coordinate axes described above is equal to $\frac{4^n}{2} - \frac{2^n}{2}$.*

**Theorem 2.** *The number of vertexes in Sierpinski gasket is equal to $3^n$ for the $n$-variable function $f$.*

**Corollary 1.** *The number of empty elements in Sierpinski gasket is equal to $\frac{4^n}{2} - \frac{2^n}{2} - 3^n$.*

**Coordinate convention 2.** Another coordinate convention assumes that each element of Sierpinski gasket has $n$-dimensional coordinates. For Boolean functions, we have to encode three states of the variable $x$: d (the variable is not present in an expression), $\overline{x}$ (variable with complement) and $x$ (variable without complement). The following values are assigned for variable's states: '1' for d, '2' for $\overline{x}$ and '3' for $x$. Such encoding has been chosen to provide the similarity to the triangle rule (Equation 1): $1 \oplus 2 \oplus 3 = 0$.

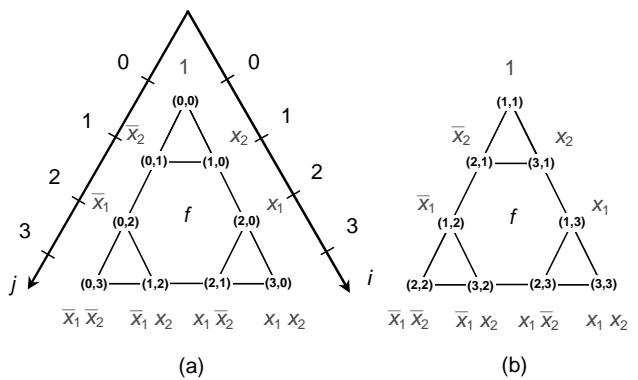

**Figure 4. Different coordinate systems**

**Theorem 3.** *The total memory size (in elements) needed to represent a $n$-variable Boolean function $f$ using Sierpinski gasket and the $n$-dimensional coordinates is $3^n$.*

These coordinate systems are shown in Figure 4. To minimize the space complexity, we have selected the second coordinate convention to manipulate with Sierpinski gasket (Boolean function) and Sierpinski pyramid (4-valued function). Recall that coordinate convention assumes that each element of Sierpinski gasket can be addressed using $n$-dimensional coordinates.

### 3.2 Applications

#### 3.2.1 Logic Expansions

Generally, a decomposition of a Boolean function $f$ with respect to arbitrary variable $x$, or, in other words, an expansion of $f$ given $x$, can be represented by the formula:

$$f = (\mathcal{C}_0 \cdot \overline{x} \cdot f_0) \circ (\mathcal{C}_1 \cdot x \cdot f_1) \circ (\mathcal{C}_d \cdot f_d), \quad (2)$$

where $C = [\mathcal{C}_0 \mathcal{C}_1 \mathcal{C}_d]$ is a vector of coefficients, and symbol $< \circ >$ denotes a logical operation, EXOR for AND/EXOR expressions. Below we will utilize the following expansions derived from possible combinations of coefficients $\mathcal{C}_0, \mathcal{C}_1, \mathcal{C}_d$: *Shannon* ($S$) expansion ($C = [110]$); *positive Davio* ($pD$) expansion ($C = [011]$), and *negative Davio* ($nD$) expansion ($C = [101]$). The expansion rules and their multiple-valued counterparts in GF(4) are given in Table 1.

The main difference between the proposed structure of Sierpinski gasket and the original one (Figure 1(a)) is that the novel structure reflects the properties of the whole set of expansions. The simple manipulations with Sierpinski gasket yield to the rules with triangles (Figure 5).

#### 3.2.2 ESOP

Sierpinski gasket contains complete information to reconstruct the function in the form of AND/EXOR expression. Thus, all coefficients of ESOP expression can be obtained from Sierpinski gasket. And vice versa, ESOP expression can be uniquely represented by Sierpinski gasket.

Each vertex of Sierpinski gasket has the equivalent in the form of logic expression (product term): (i) the ones assigned to vertexes of Sierpinski gasket correspond to product terms of ESOP expression; (ii) zeros and $d$ values are considered for manipulations only. Final expressions can be formed using EXOR operation applied for product terms. We interpret $d$ as an unspecified coefficient of ESOP expression. For arbitrary function $f$ we can build a set of Sierpinski gaskets with different costs (number of product terms, number of literals). Value $d$ allows us to manipulate with the set of gaskets, for example, to find the gasket and referred ESOP expression with the minimal cost. The manipulation rules which can lead to extension and simplification of the logic expression are presented in Figure 6(a).

**Example 4.** *The gasket from Figure 2 contains the following ones coordinates: (1,2,2), (3,1,3) and (3,3,2). These coordinates produce ESOP expression: $\overline{x}_2 \cdot \overline{x}_3 \oplus x_1 \cdot x_3 \oplus x_1 \cdot x_2 \cdot \overline{x}_3$.*

#### 3.2.3 Multiple-valued Functions

The approach that is presented above has been generalized for multiple-valued functions. Figure 7(b) gives a graphical representation of Sierpinski gasket for 4-valued function (fixed polarity $4 - pD$ in GF(4)). One can conclude that this graph denotes Sierpinski pyramid. Another pyramid of a Sierpinski kind will serve as a representation of mixed polarity expression for 4-valued function (Figure 7(c)). Manipulation rules are given in Figure 6(b). We have considered 4-valued functions, hence this approach can be generalized for functions with any radix.

**Table 1. Shannon and Davio expansions and their analogues in GF(4)**

| Type | Rule of Expansion |
|---|---|
| $S$ | $f = \overline{x} \cdot f_{|x=0} \oplus x \cdot f_{|x=1}$ |
| $pD$ | $f = f_{|x=0} \oplus x \cdot (f_{|x=0} \oplus f_{|x=1})$ |
| $nD$ | $f = f_{|x=1} \oplus \overline{x} \cdot (f_{|x=0} \oplus f_{|x=1})$ |
| $4-S$ | $f = J_0(x) \cdot f_{|x=0} + J_1(x) \cdot f_{|x=1} +$ $J_2(x) \cdot f_{|x=2} + J_3(x) \cdot f_{|x=3}$, |
| $4-pD$ | $f = f_{|x=0} + x \cdot (f_{|x=1} + 3f_{|x=2} + 2f_{|x=3})$ $+ x^2 \cdot (f_{|x=1} + 2f_{|x=2} + 3f_{|x=3})$ $+ x^3 \cdot (f_{|x=0} + f_{|x=1} + f_{|x=2} + f_{|x=3})$ |
| $1-4-nD$ | $f = f_{|x=1} + {}^{1-}x \cdot (f_{|x=0} + 2f_{|x=2} + 3f_{|x=3})$ $+ {}^{1-}x^2 \cdot (f_{|x=0} + 3f_{|x=2} + 2f_{|x=3})$ $+ {}^{1-}x^3 \cdot (f_{|x=0} + f_{|x=1} + f_{|x=2} + f_{|x=3})$ |
| $2-4-nD$ | $f = f_{|x=2} + {}^{2-}x \cdot (3f_{|x=0} + 2f_{|x=1} + f_{|x=3})$ $+ {}^{2-}x^2 \cdot (2f_{|x=0} + 3f_{|x=1} + f_{|x=3})$ $+ {}^{2-}x^3 \cdot (f_{|x=0} + f_{|x=1} + f_{|x=2} + f_{|x=3})$ |
| $3-4-nD$ | $f = f_{|x=3} + {}^{3-}x \cdot (2f_{|x=0} + 3f_{|x=1} + f_{|x=2})$ $+ {}^{3-}x^2 \cdot (3f_{|x=0} + 2f_{|x=1} + f_{|x=2})$ $+ {}^{3-}x^3 \cdot (f_{|x=0} + f_{|x=1} + f_{|x=2} + f_{|x=3})$ |

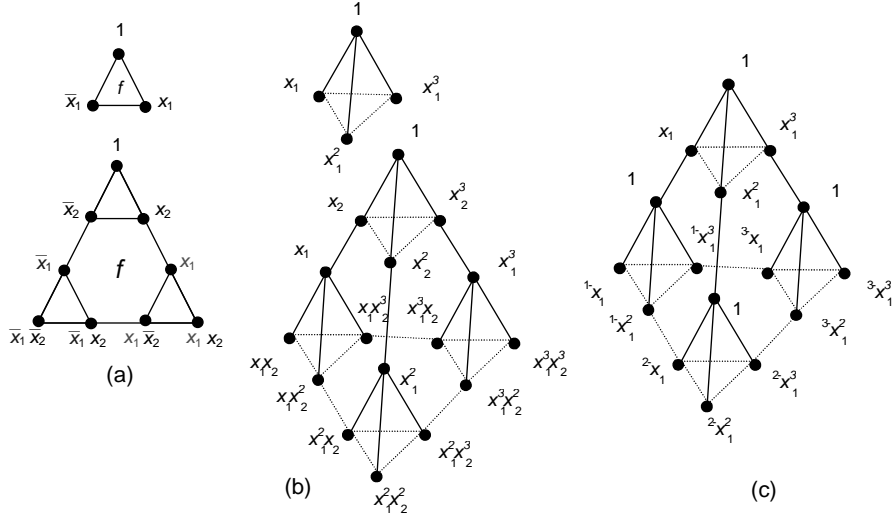

**Figure 7. The evolution of Sierpinski gasket representations for Boolean and 4-valued functions**

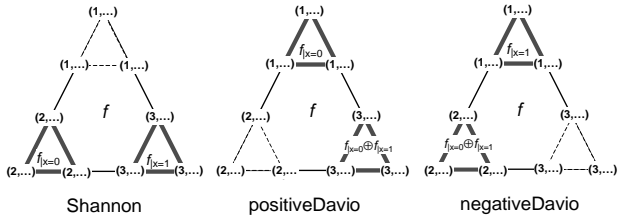

**Figure 5. Representation of expansions using Sierpinski gasket**

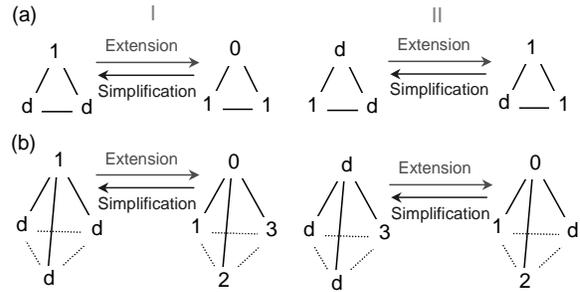

**Figure 6. Manipulation rules for 4-valued functions represented by Sierpinski gasket**

## 4 Minimization Algorithm and Experiments

The sketch of the $Sierpinski - GFSOP$ algorithm is depicted below.

### 4.1 Algorithm $Sierpinski - ESOP$ ($Sierpinski - GFSOP$)

1. Generate non-zero vertexes of Sierpinski Gasket (Sierpinski Pyramid);
2. Apply simplification or extension rules to existing Sierpinski Gasket (Sierpinski Pyramid);
3. Calculate the cost function (number of product terms or literals);
4. Choose backtracking if no improvements.

Note that this algorithm manipulates with non-zero coefficients only applying rules of fractal geometry.

**Example 5.** *The steps of ESOP minimization for the function $f = x_1 \vee \overline{x}_1 \cdot \overline{x}_2$ with truth vector $[1011]$ are given in Figure 8. Finally, the gasket contains the following ones coordinates: (2,3) and (1,1). These coordinates produce ESOP expression: $1 \oplus \overline{x}_1 x_2$.*

### 4.2 Complexity

Minimization algorithm presented here is similar to $EXORCISM - MV3$ developed in [10]. It should be mentioned that the algorithm $EXORCISM - MV3$ allows to minimize only multiple-valued input, two-valued output functions. However, in this paper, we are considering multiple-valued input, multiple-valued output functions. Our manipulation technique is based on fractal notations and graph simplifications.

The space required to store the entire Sierpinski gasket for Boolean function $f$ and ESOP minimization is $O(3^n)$ memory locations, $O(4^n)$ to represent and manipulate with fixed polarity expressions in GF(4), and $O(16^n)$ – for mixed polarity representations of 4-valued functions.

## 4.3 Experimental Results

Our $Sierpinski-ESOP$ ($Sierpinski-GFSOP$ for multiple-valued functions) program in C++ implements the above described algorithm to minimize ESOP expressions and multiple-valued functions. All the experiments have been carried out on a 800MHz Pentium PC with 128Mb of memory.

### 4.3.1 ESOP minimization

In the first series of experiments, we have selected several LGSynth93 benchmarks to build Sierpinski gaskets and minimize ESOP expressions. Thus, for xor5 with 5 inputs, the number of ones vertexes in the gasket is 5, and therefore there is 5 product terms in ESOP expression. For 9sym with 9 inputs, the number of ones vertexes is equal to 84, which is the number of product terms of ESOP expression. Table 2 contains the details of comparison between the proposed approach and two techniques: (i) symbolic manipulation EXORCISM-MV3 [10] and (ii) DD based $Est/Greedy$ [5]. In many cases, our program demonstrated superior results. Memory allocation for the selected set of benchmarks using Sierpinski gaskets is 1.4 times effective than using BDDs (for instance, 14 bytes for xor5 and 292 bytes for 9sym instead of 96 bytes and 400 bytes correspondingly in case of BDDs).

### 4.3.2 Minimization in GF(4)

In the second series of experiments, we have tested $Sierpinski-GFSOP$ on several 4-valued benchmarks (Table 3). The 4-valued benchmarks were generated by pairing inputs and outputs of MCNC benchmarks. The experimental results demonstrate that $Sierpinski-GFSOP$ produces the fewer number of terms and literals against the program based on information theoretic measures $Info-MV^{PSDRMGF}$ (see [13] for details), however the program $Info-MV$ is extremely faster.

## 5 Concluding Remarks

We have presented a novel approach for representing logic functions using fractals, namely the structures of Sierpinski gasket and Sierpinski pyramid are used. Several related issues such as coordinate convention and memory allocation have been discussed. Due to the recursive nature of Sierpinski gasket, the proposed techniques for constructing triangles are efficient in both run-time and storage. It becomes possible to develop manipulation algorithms, like ESOP minimization or minimization of multiple-valued functions in GF(4), using representation in the form of Sierpinski gasket.

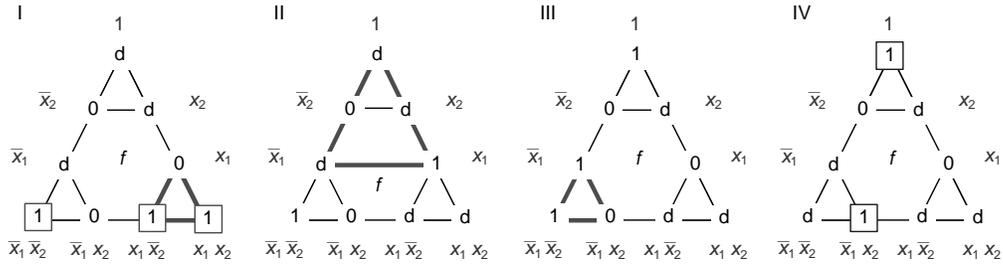

**Figure 8. Steps of ESOP minimization using Sierpinski gasket**

**Table 2. Experimental results in ESOP minimization**

|  | **I/O** | $EXORCISM-MV3$ [10] $T/L$ | $Est/Greedy$ [5] $T/L$ | $Sierpinski-ESOP$ $T/L$ |
|---|---|---|---|---|
| xor5 | 5/1 | 5/10 | - | 5/**5** |
| rd53 | 5/3 | **14**/57 | - | 17/**48** |
| bw | 5/28 | **22**/319 | 20/- | 26/**122** |
| inc | 7/9 | **26**/176 | 31/- | 32/**153** |
| 5xp1 | 7/10 | **32**/170 | 47/- | 55/**166** |
| 9sym | 9/1 | **51/426** | - | 84/449 |
| apex4 | 9/19 | 439/6181 | - | **438**/**3692** |
| sao2 | 10/4 | 28/288 | 41/- | 55/370 |
| ex1010 | 10/10 | **670**/7466 | - | 879/**7014** |
| duke2 | 22/29 | **78**/909 | 108/- | 91/**793** |
| x6dn | 39/5 | - | 104/- | 113/1026 |
| x7dn | 66/15 | - | - | 594/4610 |

**Table 3. Comparison of $Info-MV$ [13] and $Sierpinski-GFSOP$ on 4-valued benchmarks**

|  | **I/O** | $Info\text{--}MV^{PSDRMGF}$ $T/L/t$ | $Info\text{--}MV^{PSDKGF}$ $T/L/t$ | $Sierpinski-GFSOP$ $T/L/t$ |
|---|---|---|---|---|
| 5xp1 | 3/5 | 165/521/0.04 | 142/**448**/0.39 | **140**/453/4.53 |
| 9sym | 5/1 | 246/800/0.12 | 201/776/0.29 | **193**/**764**/30.87 |
| bw | 3/14 | 54/144/0.02 | 44/**132**/0.02 | **42**/140/5.91 |
| clip | 5/3 | 825/3435/4.50 | 664/2935/4.51 | **606**/**2802**/40.03 |
| con1 | 4/1 | 50/138/0.53 | 19/50/0.42 | **18**/**48**/12.75 |
| ex1010 | 5/5 | 1018/4076/8.06 | 997/4985/8.61 | **912**/**3977**/145.80 |
| inc | 4/5 | 146/493/0.36 | 65/216/0.82 | **60**/**198**/15.49 |
| misex1 | 4/4 | 48/108/3.6 | 15/**38**/0.02 | **15**/43/9.76 |
| rd84 | 4/2 | 207/656/0.85 | 207/656/1.05 | 207/656/9.96 |
| sao2 | 5/2 | 252/1133/4.8 | 96/437/0.24 | **81**/**365**/37.06 |
| squar5 | 3/4 | 51/135/0.01 | 48/128/0.03 | **42**/**112**/10.15 |

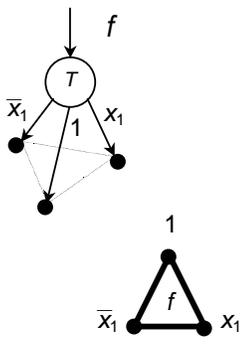
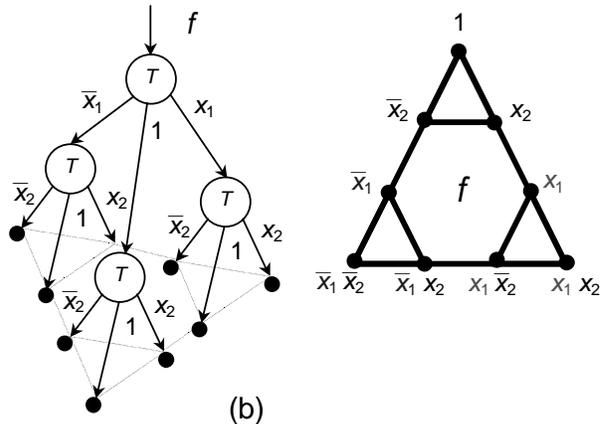

(a)

(b)